\documentclass[a4paper,12pt]{article}
\usepackage{epsfig}
\date{\today}  
\newcommand{\bmat}{\left(\begin{array}}
\newcommand{\emat}{\end{array}\right)}
\def\NPB#1#2#3{Nucl. Phys. B{#1} (19#2) #3}
\def\PLB#1#2#3{Phys. Lett. B{#1} (19#2) #3}

\def\PRD#1#2#3{Phys. Rev. D{#1} (19#2) #3}
\def\PRL#1#2#3{Phys. Rev. Lett. {#1} (19#2) #3}

\def\lsim{\raise0.3ex\hbox{$\;<$\kern-0.75em\raise-1.1ex\hbox{$\sim\;$}}}
\def\gsim{\raise0.3ex\hbox{$\;>$\kern-0.75em\raise-1.1ex\hbox{$\sim\;$}}}

\def\yzero{\smash{\hbox{$y\kern-4pt\raise1pt\hbox{${}^\circ$}$}}}

\def\-{\hphantom{-}}

\def\s2{\frac{1}{\sqrt2}}

\def\beq{\begin{equation}}
\def\eeq{\end{equation}}
\def\beqa{\begin{eqnarray}}
\def\eeqa{\end{eqnarray}}

\def\IF{\relax{\rm I\kern-.18em F}}
\def\II{\relax{\rm I\kern-.18em I}}
\def\IP{\relax{\rm I\kern-.18em P}}
\def\IC{\relax\hbox{\kern.25em$\inbar\kern-.3em{\rm C}$}}
\def\IR{\relax{\rm I\kern-.18em R}}

\def\Dsl{\,\raise.15ex\hbox{/}\mkern-13.5mu D} 
\def\IZ{Z\kern-.4em  Z}
\def\bmat{\left(\begin{array}}
\def\emat{\end{array}\right)}

\def\amu{a_{\mu}}
\def\SM{ {\rm SM}}
\def\mch{m_{\chi^{\pm}}}
\def\mneut{m_{\chi^{0}}}
\def\tgb{\tan{\beta}}

\def\asusy{$a_{\mu}^{{\rm \small SUSY}}$}

\def    \part          {\partial}
\def    \be            {\begin{equation}}
\def    \ee            {\end{equation}}
\def    \bea           {\begin{eqnarray}}
\def    \eea           {\end{eqnarray}}
\def    \nn            {\nonumber}

\begin{document}
%
\pagestyle{empty}
\rightline{FTUAM 01/07}
\rightline{IFT-UAM/CSIC-01-13}
\rightline{HIP-2001-08/TH}
\rightline{SUSX-TH/01-017}
\rightline{April 2001}

\renewcommand{\thefootnote}{\fnsymbol{footnote}}
\setcounter{footnote}{0}

\vspace{0.0cm}
\begin{center}
\large{\bf Muon anomalous magnetic moment 
in supersymmetric scenarios with an intermediate scale
and nonuniversality\\[5mm]}
\mbox{
\sc{
\small
{D.G. Cerde\~no$^{1,2}$,
E. Gabrielli$^{3}$,
S. Khalil$^{4,5}$,
C. Mu\~noz$^{1,2}$,
E. Torrente-Lujan$^{1}$
}
}
}
\begin{center}
{\small
{\it $^1$ Departamento de F\'{\i}sica
Te\'orica C-XI, Universidad Aut\'onoma de Madrid,\\[-0.1cm]
Cantoblanco, 28049 Madrid, Spain. \\
\vspace*{2mm}
\it $^2$ Instituto de F\'{\i}sica Te\'orica  C-XVI,
Universidad Aut\'onoma de Madrid,\\[-0.1cm]
Cantoblanco, 28049 Madrid, Spain.\\
\vspace*{2mm}
\it $^3$ Helsinki Institute of Physics,
P.O. Box 64, FIN-00014 Helsinki, Finland.\\
\vspace*{2mm}
\it $^4$ Centre for Theoretical Physics, University of Sussex, 
Brighton BN1 9QJ, U.K.\\
\vspace*{2mm}
\it $^5$ Ain Shams University, Faculty of Science, Cairo
11566, Egypt.} 
}
\end{center}

{\bf Abstract} 
\\[7mm]
\end{center}
\begin{center}
\begin{minipage}[h]{15.0cm}

We analyze the anomalous magnetic moment of the muon $a_{\mu}$ in 
supersymmetric scenarios. First we concentrate on scenarios with 
universal soft terms. 
We find that a moderate increase of $a_{\mu}$
can be obtained by lowering the unification 
scale $M_{GUT}$ to intermediate values $10^{10-12}$ GeV. 
However, large values of $\tan \beta$ are still
favored. 
Then we study the case of non-universal soft terms.
For the usual value $M_{GUT}\approx 10^{16}$ GeV, we 
obtain $a_{\mu}$ in the favored experimental range  
even for moderate $\tan\beta$ regions 
($\tan\beta\gsim 5$).
Finally, we give an explicit example of these scenarios.
In particular, we show that in a D-brane model, 
where the string scale is naturally of order $10^{10-12}$ GeV and 
the soft terms are non universal, $a_{\mu}$ is enhanced with low $\tan\beta$.

\end{minipage}
\end{center}

\vspace{0.4cm}
\begin{center}
\begin{minipage}[h]{14.0cm}
PACS: 14.60.Ef, 12.60.Jv, 11.25.Mj

Keywords: muon magnetic moment, scales, supersymmetry, superstrings.
\end{minipage}
\end{center}
\newpage
\setcounter{footnote}{0}
\setcounter{page}{1}
\pagestyle{plain}

%
\section{Introduction}

Recently, an intense theoretical activity about new physics contributions
to the anomalous magnetic moment of the muon ($\amu$) 
has appeared in the literature \cite{g2_rev}--\cite{nath1}.
This has been motivated by
the new measurement in the E821 experiment 
at the Brookhaven National Laboratory (BNL) \cite{BNL}, where
a 2.6 $\sigma$ deviation from the standard model (SM) predictions 
\cite{g2_rev,davier} was reported
\beq
\amu ({\rm E821}) -\amu (\SM) = (43 \pm 16) \times 10^{-10}\ .
\label{bnl}
\eeq

There are also criticisms about the SM prediction quoted by this experiment
\cite{yndurain}\footnote{It is worth noticing that a recent 
paper \cite{marro} refutes these arguments.}.
It is not impossible that this deviation is due to the
hadronic contribution to the vacuum polarization \cite{davier}, which is
the largest source of error reflecting the large 
experimental uncertainty of the data.
However, the possibility that new physics effects are
at the origin of the BNL deviation is very exciting.
First of all, the new physics scale should be pretty close 
to the electroweak threshold, since its contributions to $\amu$ are
of the same order or even larger than the 
corresponding electroweak corrections \cite{g2_rev}. Besides, it
has to pass all the electroweak precision tests of the SM and
must be in agreement with all the known 
results from accelerators experiments.

The minimal supersymmetric standard model (MSSM) 
is a well established candidate for such a theory\footnote{see 
ref.\cite{g2_exot}
for alternative 
possibilities which might explain this deviation, such as 
leptoquarks, compositness, large extra dimensions models, etc.}.
On a general ground, if the MSSM is responsible
of the BNL deviation, then the supersymmetric (SUSY) particle spectrum
should be in the expected discovery range of  
Fermilab 2 TeV $p\bar{p}$ 
collider and certainly of the Large Hadronic $p p$ Collider (LHC) 
at CERN.

The most popular SUSY-breaking scenarios in the MSSM have been recently
reanalyzed in the light of the new BNL results \cite{g2_new}.
In particular, in the supergravity scenario the 
main conclusions can be summarized as follows.
The requirement that the SUSY contribution to $\amu$ is 
within the 2 $\sigma$ level in eq.(\ref{bnl}), 
leads to the following 
upper limits on the lightest chargino and neutralino mass, respectively
$\mch \lsim 600$ GeV and $\mneut \lsim 300$ GeV 
for $\tgb \leq 30$. The corresponding upper bound on 
the sneutrino masses is weaker and of the order
of 1 TeV \cite{nath1}.

The above analysis was performed assuming universality 
of the soft-breaking terms at the unification scale, 
$M_{GUT} \approx 10^{16}$ GeV, as is usually done in the MSSM literature.
As known, such a scale can be obtained 
in the superstring framework, in particular this is the case 
of weakly coupled heterotic string \cite{reviewdienes}, 
type I string \cite{Witten,Rigolin} and 
heterotic M-theory  \cite{Witten,Banks}.
However, recently, it was realized that 
the string scale may be anywhere between the weak and the Plank scale.
For instance, D-brane configurations where the SM lives, allow these
possibilities in type I strings 
\cite{Lykken}-\cite{nosotros}, 
and similar results can also be
obtained in type II strings \cite{typeII} 
as well as in strongly and weakly coupled 
heterotic strings \cite{stronghete,weakandstronghete}.

To use the value of 
the initial scale, say $M_I$, as a free parameter 
for the running of the soft terms
is particularly interesting since 
there are several arguments in favor of SUSY scenarios
with scales 
$M_I\approx 10^{10-14}$ GeV.
First, these scales were suggested in \cite{stronghete}
to explain many experimental observations as
neutrino masses or the scale for axion physics. 
Second, with the string scale of order 
$10^{10-12}$ GeV
one is able
to attack the hierarchy problem of unified theories 
without invoking any hierarchically suppressed non-perturbative
effect \cite{typeIinter}. Third,
for intermediate scale scenarios, 
charge and color breaking constraints become less important \cite{Allanach}.
Let us recall that, due to these constraints,
when working with the usual
unification scale, $M_{GUT}\approx 10^{16}$ GeV, 
there are extensive regions in the
parameter
space of soft SUSY-breaking terms that become
forbidden \cite{mua}.
There are other arguments in favor 
of scenarios with initial scales $M_I$ smaller
than $M_{GUT}$.
For example these scales
might also
explain the observed ultra-high energy ($\approx 10^{20}$ eV) cosmic rays
as products of long-lived massive string mode decays. Besides,
several models of chaotic inflation favor also these scales \cite{caos}.
Finally, D-brane models lead naturally to intermediate values for
the string scale, in order to reproduce low-energy data \cite{nosotros}.

Inspired by these scenarios, 
it was recently pointed out that the
neutralino-nucleon cross sections, which are relevant
for dark matter experiments,  
are very sensitive to the variation of the initial scales 
for the running of the soft SUSY--breaking terms \cite{muas,bailin,nosotros}.
In particular, it was found that the smaller the scale is the larger the 
cross sections become. For instance, by taking $10^{10-12}$ GeV rather than 
$M_{GUT}$, extensive regions in the parameter space of the MSSM 
have been found \cite{muas}
where the neutralino-nucleon cross sections
are in the expected range of sensitivity of 
DAMA \cite{experimento1} and CDMS \cite{experimento2} detectors, and this
even for moderate $\tgb$ regions  ($\tgb \geq 3$). This analysis 
was performed in the universal scenario for the soft terms. 
In contrast, in the usual case with initial scale at $M_{GUT}$,
these large cross sections are achieved only for $\tgb > 20$ 
\cite{Bottino,Gomez}.

The fact that smaller initial scales imply larger 
neutralino--nucleon cross sections can be basically understood as follows.
These cross sections are very sensitive to the $\mu$ parameter, 
which is the standard coupling in the superpotential 
between the two Higgs doublets, since 
they increase when  
$\mu$ decreases. Furthermore, the value of $\mu$ is also very sensitive 
to the initial scale $M_I$ and it decreases when $M_I$
decreases. As a consequence, decreasing $M_I$ one obtains
larger cross sections \cite{muas}.

One of the purposes of the present paper is to analyze, 
in the light of the new BNL results and in connection to the work
of ref.\cite{muas}, the variations of $\amu$ as a 
function of the initial scale $M_I$.
The main reason is that in the MSSM $\amu$ is expected to be 
particularly sensitive to the $\mu$ parameter and therefore to the 
initial scale.

On the other hand, the soft SUSY-breaking terms can have in general
a non-universal structure in the MSSM. Such non-universality
can be derived from supergravity and superstring models \cite{dilaton}.
In fact, it was shown 
in ref.\cite{Bottino} that non-universal scenarios
allow for a remarkable 
enhancement of the neutralino-nucleon cross section to be in the current 
experimental regions, and this even for $\tan\beta >4$. 
Here and along this line, we will analyze 
the effect induced on \asusy\ by the  non--universality of the soft terms. 

Finally, 
we give an explicit example where both situations, non-universal soft
terms and an intermediate scale, 
are realized. This is the case of a D-brane model.

The paper is organized as follows. In section 2 we review general formulae
for the SUSY contributions to $a_{\mu}$. In section 3 we study the 
prediction for \asusy\ in SUSY scenarios with universal
soft terms, when intermediate scales are allowed.
Section 4 is devoted to the study of the effect of the non--universality 
of the soft terms on \asusy. This is carried out
first  in the context of the MSSM with the usual scale
$M_{GUT}\approx 10^{16}$ GeV, and second
in the framework of D--brane constructions. 
The conclusions are given in section 5. 

\section{SUSY contributions to the muon anomalous magnetic moment}
The supersymmetric contributions to $\amu$ are mainly via magnetic--dipole
penguin diagrams with an exchange of sneutrino--chargino or 
smuon--neutralino in the loop. These contributions can be found
in the literature \cite{nath}-\cite{g2_gs}, and they are given by: 
\begin{eqnarray}
a_{\mu}^{\chi^0} & = & \frac{m_{\mu}}{16 \pi^2} \sum_{m,i}
\left\{
- \frac{m_{\mu}}{6 m_{\tilde{\mu}_m}^2 \left(1 - x_{mi}\right)^{4}}   
\left(\vert N_{mi}^L \vert^2 + \vert N_{mi}^R \vert^2 \right)
\right.
\nonumber\\   
& \times & \left( 1 - 6 x_{mi} + 3 x^2_{mi} + 2 x_{mi}^3 - 6 x_{mi}^2 
\ln x_{mi} \right)
\nonumber\\
& + &
\left.
\frac{m_{\chi^0_i}}{m_{\tilde{\mu}_m}^2 ( 1 - x_{mi})^{3}}
N_{mi}^L N_{mi}^R ( 1 - x_{mi}^2 + 2 x_{mi} \ln x_{mi}) \right\}\ ,
\label{neutralino}
\end{eqnarray}
\begin{eqnarray}
a_{\mu}^{\chi^{\pm}} & = & \frac{m_{\mu}}{16 \pi^2} \sum_k \left\{ 
\frac{m_{\mu}}{3 m_{\tilde{\nu}}^2 \left(1-x_k\right)^4}
\left(\vert C_k^L \vert^2 + \vert C_k^R \vert^2 \right)
\right.
\nonumber\\
& \times & \left( 1 + 1.5 x_k + 0.5 x_k^3
- 3 x_k^2 + 3 x_k \ln x_k \right)
\nonumber\\
& - & 
\left.
\frac{ 3 m_{\chi^\pm_k}}{m_{\tilde{\nu}}^2 \left(1-x_k\right)^3} C_k^L
C_k^R \left( 1 - \frac{4 x_k}{3} + \frac{x_k^2}{3} + \frac{2}{3}
\ln x_k\right) \right\}\ ,
\label{chargino}
\end{eqnarray}
where $x_{mi} = m_{\chi^0_i}^2/m_{\tilde{\mu}_m}^2$, $x_k = m_{\chi^\pm_k}^2/
m_{\tilde{\nu}}^2$\ ,
\begin{eqnarray}
N_{mi}^L & = &  h_{\mu}
(U_{\chi^0})_{3i} 
(U_{\tilde{\mu}})_{Lm}
+ \sqrt{2} g_1 (U_{\chi^0})_{1i} (U_{\tilde{\mu}})_{Rm}\ ,
\nonumber\\
N_{mi}^R & = & - h_{\mu}
(U_{\chi^0})_{3i} 
(U_{\tilde{\mu}})_{Rm}
+ \frac{g_2}{\sqrt{2}}  (U_{\chi^0})_{2i} (U_{\tilde{\mu}})_{Lm} +
\frac{g_1}{\sqrt{2}} (U_{\chi^0})_{1i} (U_{\tilde{\mu}})_{Lm}\ ,
\nonumber\\
C_k^L & = & h_{\mu}
U_{k2}\ ,
\nonumber\\
C_k^R & = & - g_2 V_{k1}\ .
\label{vertex}
\end{eqnarray}
Here $(U_{\chi^0})_{ij}$ with $i,j=1,4$, $(U_{\tilde{\mu}})_{(R,L)m}$
with $m=1,2$, and $U_{kl}$, $V_{kl}$ with $k,l=1,2$
are the neutralino, smuon and chargino mixing matrices respectively,
$m_{\chi^0_i}$, $m_{\tilde{\mu}_m}$, $m_{\tilde{\nu}}$ and $m_{\chi^\pm_k}$
are the neutralino, smuon, sneutrino and chargino mass
eigenstates respectively, $m_{\mu}$ is the muon mass,
$h_{\mu}$ is the Yukawa coupling of the muon and $g_i$ are
the electroweak gauge couplings. 
Let us remark that we are using the following sign conventions for Yukawa
couplings, and gaugino and Higgsino masses in the Lagrangian:
${\cal L}=-h_u H_u^0 \bar u_L u_R -h_d H_d^0 \bar d_L d_R 
- h_e H_d^0 \bar e_L e_R +
\frac{1}{2}\sum_{a} M_a {\lambda}_a \lambda_a + 
\mu \tilde{H}^0_u \tilde{H}^0_d$ + h.c., with
the neutralino basis given by
($\tilde B^0=-i{\lambda}'$, $\tilde W_3^0=-i{\lambda}^3$, $\tilde H_u^0$, 
$\tilde H_d^0$). 

Eqs.(\ref{neutralino}-\ref{vertex}) show that the dominant contributions
to $a_{\mu}^{\chi^{\pm}}$ and $a_{\mu}^{\chi^0}$
correspond to the terms with $C^LC^R$ and $N^LN^R$ \cite{moroi}, since
they are proportional to the chargino and neutralino masses respectively.
Furthermore, it was found that the chargino contribution
dominates the neutralino contribution \cite{moroi}.
Note e.g. that the lightest neutralino 
$\chi_1^0$
is often bino--like, i.e.
$(U_{\chi^0})_{11} \sim 1$ and $(U_{\chi^0})_{1i} <<1$ for $i=2,3,4$. 
Therefore
the terms proportional to  $g_1^2$ 
are expected to give the dominant contribution to $a_{\mu}^{\chi^0}$. 
However, these terms are always suppressed by the matrix entries
$(U_{\tilde{\mu}})_{L2}$ or $(U_{\tilde{\mu}})_{R1}$
which are of order $m_{\mu}/m_{SUSY}$.
On the contrary, the chargino contribution $C^LC^R$
do not have such a suppression in the chargino mixing.

Eqs.(\ref{neutralino}-\ref{vertex}) also show that \asusy\ becomes larger as 
$\tan \beta$, the ratio of Higgs vacuum expectation values
$\langle H_2\rangle/\langle H_2\rangle$,
increases \cite{tan,nath}. Recall in this sense that 
for $\tan \beta$ not too small the dominant 
chargino 
contribution can be approximated as \cite{g2_gs}
\beq
a_{\mu}^{\chi^{\pm}}\approx \frac{3\alpha_2}{4\pi}\tgb
\frac{m_{\mu}^2 \mu M_2}{m^2_{\tilde{\nu}} (M_2^2 - \mu^2)} \left[ f(x_{M_2}) -
f(x_{\mu}) \right]\ ,
\label{loopFs}
\eeq
where
$x_{M_2}=M_2^2/m^2_{\tilde{\nu}}\ ,\ x_{\mu}=\mu^2/m^2_{\tilde{\nu}}$\ ,
$M_2$ is the weak gaugino mass and $f$ is a loop function defined as
\beq
f(x)=\frac{3-4x+x^2+2\log(x)}{3(1-x)^3}\ .
\label{loopF}
\eeq

This approximate formula helps also to draw some important conclusions on the 
SUSY contributions to $a_{\mu}$, 
which we have checked that are 
still valid for low $\tan \beta$ as well. 

First, decreasing the values of $M_2$, $\mu$ and $m^2_{\tilde{\nu}}$ leads 
to increase \asusy. Indeed, different scenarios that enhance \asusy\ (as 
we will show in next sections) are based on the decrease of these quantities.

Second, The sign of \asusy\ is given by the sign of the product
$\mu M_2$ since the factor 
$(f(x_{M_2}) - f(x_{\mu}))/(M_2^2 - \mu^2)$
is positive in general. 
Assuming $M_2$ is real and positive (after performing $U(1)_R$ rotation), the 
positiveness of \asusy\ 
implies that $\mu$ should be positive. This has interesting 
consequences for the $b\to s \gamma$ constraints (at large $\tan \beta$) and 
also for the dark matter detection rate. It is known that for $\mu <0$ 
the neutralino-nucleon cross section is reduced a lot due to accidental 
cancellations between different contributions. Also experimental
constraints coming from the $b\to s \gamma$ process highly reduce
the $\mu <0$ parameter space.

Third, as it is known, in order to satisfy the Higgs mass 
bound ($m_{H} \gsim 114$ GeV)
large stop masses are required to increase the radiative corrections to
the 
Higgs
mass. 
However, as mentioned above, large values of \asusy\ require light sneutrino 
and smuon. Thus a 
non--universal pattern of the soft SUSY-breaking terms would be 
preferred to fulfil both conditions. 
In particular, a pattern with light sleptons and 
heavy squarks. 

Fourth, the trilinear coupling $A$ appears in left--right smuon mixing,
following our sign conventions discussed below eq.(\ref{vertex}), as 
$m_{\mu}(A-\mu \tan \beta)$ and stop mixing as $m_{top}(A-\mu \cot \beta)$.
It has 
a significant effect on the stop mass and large values of $\vert A \vert$ are 
favored. 
On the contrary, 
the smuon mixing is dominated by the $\mu$ term (specially in the large 
$\tan \beta$ region). However, for 
low $\tan \beta$ (and also low $\mu$) 
$A$-terms could have important effects if 
$A \simeq \mathcal{O}(-3m)$.

\section{$\amu$ in SUSY models with intermediate scale}

In this section we consider the predictions for $\amu$ in the MSSM as a 
function of the initial scale $M_I$ for the soft SUSY--breaking terms.
As discussed in the Introduction, there are very interesting arguments
in favour of scenarios with scales $M_I$ smaller than $M_{GUT}$.
Following the analysis of ref.\cite{muas} we will consider 
two possible scenarios with ``intermediate'' initial scales, concerning
the unification of the gauge couplings.

First, we will assume that
these are non universal and their values
will depend on the initial scale $M_I$ chosen. 
For instance, for the scale $M_I=10^{11}$ GeV,
one obtains $g_3\approx 0.8$, $g_2\approx 0.6$ and $g_1\approx 0.5$.
This scenario might be inspired for example by 
D-brane configurations where the SM lives.
If the SM comes from the
same collection of D-branes, 
stringy corrections might change the boundary conditions at the string scale
$M_I$ in order to mimic the effect of field theoretical logarithmic running 
\cite{Rigolin,mirage}. 
Another possibility giving rise to a similar result
might arise when
the gauge groups came from different types of D-branes.
Since different D-branes have associated different couplings,
this would imply the non universality of the gauge couplings
(see ref.\cite{nosotros} and references therein). 

On the other hand, to obtain gauge coupling 
unification at $M_I$, $\alpha_i=\alpha$, is possible
with  
the addition of extra fields
in the massless spectrum \cite{typeIinter}.
An example of additional particles which can produce the
beta functions, $b_3=-3$, $b_2=3$, $b_1=19$, 
yielding unification at around $M_I=10^{11}$ GeV
was given in ref.\cite{Allanach},
$ 2\times [(1,2,1/2)+(1,2,-1/2)]
+ 3\times [(1,1,1)+(1,1,-1)]$,
where the fields transform under the SM gauge group.
In this example one has $g(M_I)\approx 0.8$. 

It was obtained in ref.\cite{muas}
that, due to the different values of
the gauge couplings at $M_I$, the above scenarios give rise 
to qualitatively different results for neutralino--nucleon cross sections. 
In this section we will also analyze this issue for $a_{\mu}$.

Let us concentrate first on the scenario with non-universal gauge couplings
at $M_I$. We assume, as in the minimal supergravity scenario, universality in 
the soft-breaking sector.
As usual, we eliminate the free parameter $\mu$ which appears in the 
superpotential $W = -\mu H_1^0 H_2^0$, by requiring the correct electroweak 
breaking at the $M_Z$ scale.
These requirements leave us with the following 
independent parameters at the initial scale $M_I$:
$m,~M_{1/2},~A,~\tgb$, 
and the sign $(\mu)$, respectively the common scalar mass,
gaugino mass, the coefficient of trilinear terms, and the ratio of Higgs 
vacuum expectation values. 
Let us finally remark that we are assuming gaugino mass universality
at the high energy scale, 
although in this scenario gauge couplings do not unify. 
This situation is in principle
possible in generic supersymmetric models, however it is not so natural in
supersymmetric models from supergravity 
where gaugino masses and gauge couplings
are related through the gauge kinetic function. 
Since an explicit string construction with nonuniversal
gauge couplings and gaugino masses will be analyzed in detail
in Subsection~4.2, we choose to simplify the discussion here
assuming gaugino mass universality.

As emphasized in 
ref.\cite{muas}, lowering the unification scale decreases the 
value of $\mu$. Thus we will analyze here what is the influence
of this decrease on
\asusy. Let us write eq.(\ref{loopFs}) as
\beq
a_{\mu}^{\chi^{\pm}}\approx \frac{3\alpha_2}{4\pi}\tgb\ 
m_{\mu}^2\ x_{\mu}^{1/2}\ x_{M_2}^{1/2}\  
F(x_{M_2}, x_{\mu}) 
\ ,
\eeq
where 
$F(x_{M_2}, x_{\mu}) = (f(x_{M_2}) -f(x_{\mu}))/(M_2^2 -\mu^2)$
is a function which depends on $\mu$, $M_2$ and $m_{\tilde\nu}$.
It turns out that when we lower the scale, the variation of 
$\mu$ is much more important than the variation of $M_2$ and 
$m_{\tilde\nu}$. Although this produces an important decrease
in $x_{\mu}$ (while the increase in $x_{M_2}$ is moderate),
the big increase in $F$ compensates it.
In this way, higher values of \asusy\ can be obtained.

\begin{figure}[t]
\begin{center}
\psfig{file=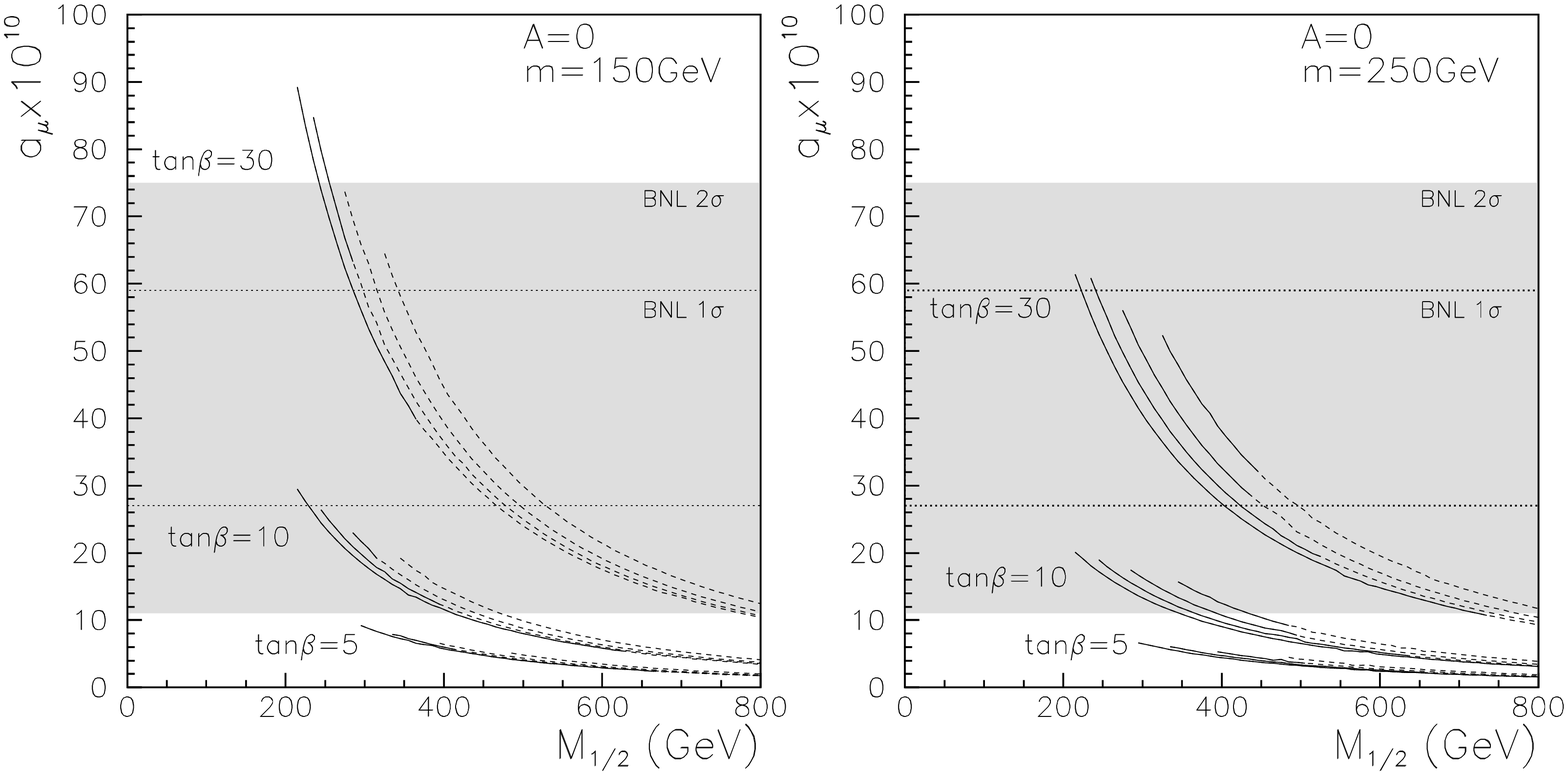, height=7cm, width=15cm}
\end{center}
\vspace{-0.5cm}
\caption{ \asusy\ as a function of the common gaugino mass $M_{1/2}$.
The four curves inside each set associated
to a particular value of $\tan\beta$ correspond, from bottom to top, to
$M_I=10^{16,14,12,10}$ GeV respectively.
Continuous lines correspond to regions where the neutralino is the LSP.}
\label{fig_150}
\end{figure}

We recall that low initial scales
play a crucial role in increasing the spin-independent part of the
neutralino-nucleon cross sections, mainly due to the decrease of
the $\mu$ parameter \cite{muas}.
In the MSSM with universal scenario at $M_{GUT}$ these 
cross sections are strongly suppressed due to the fact that 
the lightest neutralino is mainly Bino. 
By decreasing the value of the $\mu$
parameter, the Higgsino components of the lightest neutralino increase
and therefore also the spin-independent part of the cross sections increases.
On the contrary, the sensitivity of \asusy\ versus the initial scale
is quite moderate. 

\begin{figure}[t]
\begin{center}
\psfig{file=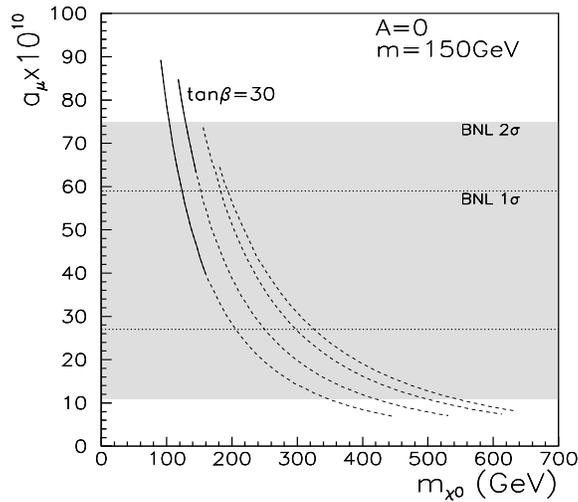, width=8cm, height=7cm
}
\end{center}   
\caption{\asusy\ as a function of the lightest neutralino mass $m_{\chi^0}$ .
The four
curves correspond, from bottom to top, to $M_I=10^{16,14,12,10}$ GeV
respectively.}
\label{a-lsp}
\end{figure}

We show the results of our analysis in  Figs.~\ref{fig_150} and \ref{a-lsp}.
They have been obtained using the general formulae 
(\ref{neutralino}-\ref{vertex}) discussed in 
Section 2.
These figures correspond to the $\mu >0$ case. 
We have not included the scenarios with opposite values of $\mu$ since they 
imply negative values 
for \asusy\ and therefore are ruled out by the BNL results.

In Fig.~\ref{fig_150} we plot \asusy\ versus the common gaugino mass at the 
initial scale, $M_{1/2}$, for a fixed value of $m=150, 250$ GeV, and $A=0$.
Inside each plot there are three sets of four curves which correspond
to $\tgb =5, 10, 30$.
The four curves inside each set correspond to 
$M_I=10^{16, 14, 12, 10}$ GeV,
from bottom to top respectively, and the 
continuous lines correspond to regions where the neutralino is the lightest
SUSY particle (LSP). Finally,
the large grey area stands for the BNL deviation at $2\sigma$ level. 
We have checked that our results are consistent
with present bounds coming from accelerators. These are LEP and 
Tevatron bounds on supersymmetric masses and CLEO 
$b\rightarrow s\gamma$ branching ratio measurements.
The former are the reason why regions with $M_{1/2}\lsim 200$ GeV 
are not allowed. In particular, 
the Higgs mass 
bound ($m_{H} \gsim 114$ GeV) is not fulfilled.
Although not shown in the figure, $b\rightarrow s\gamma$
results constrain the value of \asusy $\times 10^{10}$ to be smaller
than about 60(45) for $\tan\beta = 30$ in the case $m=150(250)$ GeV.


As shown in Fig.~\ref{fig_150}, for a given $\tgb$
the smaller the scale is, the larger \asusy\ becomes.
We have checked that the dominant contribution is due to the chargino,
as discussed in the previous section. For example, for
$\tan\beta=30$, the neutralino
contribution is not only small but also
decreases going from $M_{GUT}$ to $10^{10}$ GeV and becomes negative.
In any case, as mentioned above, 
the sensitivity of \asusy\ to the scale is quite moderate.

On the other hand, as discussed in the previous section, we obtain that
\asusy\ increases with $\tgb$.
Besides, also the deviation with the scale in \asusy\ increases with $\tgb$.
By comparing the plots with $m=150$ GeV and $m=250$ GeV
we see that \asusy\ decreases when $m$ increases, 
due to the fact that the smuon and sneutrino become heavier, but for fixed 
$\tgb$ the scale dependence in \asusy\ remains essentially the same.

Thus
the main conclusion drawn from the results in Fig.~\ref{fig_150} is
that, 
within
the $2\sigma$ level of the BNL deviation, the low $\tgb$ regions 
(namely $\tgb \le 5$) are excluded for any scale in the range
$M_I=10^{10-16}$ GeV. Besides, the sensitivity to the scale is quite 
moderate even for large $\tgb$.
Note however that for $M_{1/2}$ between 350 and 450 GeV
and $\tan\beta=10$ whereas the value of $a_{\mu}$ is in the
forbidden region for $M_{GUT}$, it is in the allowed region for
intermediate scales.

The plots in Fig.~\ref{fig_150} have been obtained for $A=0$.
However, we have checked that \asusy\ is quite insensitive to this choice, 
in particular, taking $A= \pm M_{1/2}$ 
the corresponding results are slightly modified.

In Fig.~\ref{a-lsp} we show the dependence of \asusy\ versus the lightest  
neutralino mass
varying the scale in the
same range of Fig.~\ref{fig_150}, for the representative case of 
$\tgb=30$ and $m=150$ GeV.
By requiring that \asusy\ is within the lower bound of the $2\sigma$ BNL 
region in Fig.~\ref{a-lsp}, the following upper bounds
for the SUSY mass 
can be obtained: 
$m_{\chi^0} \le 340(540)$ GeV
with 
$M_I=10^{16}(10^{10})$ GeV.

The same analysis can be carried out for 
the lightest chargino, smuon and 
sneutrino with the result
$m_{\chi^{\pm}} \le 640(560)$ GeV, 
~~$m_{\tilde{\mu}} \le 330(300) $ GeV,~ 
and $m_{\tilde{\nu}} \le 560(460)$ GeV.
Smaller (and therefore less conservative) upper bounds 
can be obtained by taking smaller values of $\tgb$ and/or larger values
of the common scalar mass $m$.

\begin{figure}
\begin{center}
\psfig{file=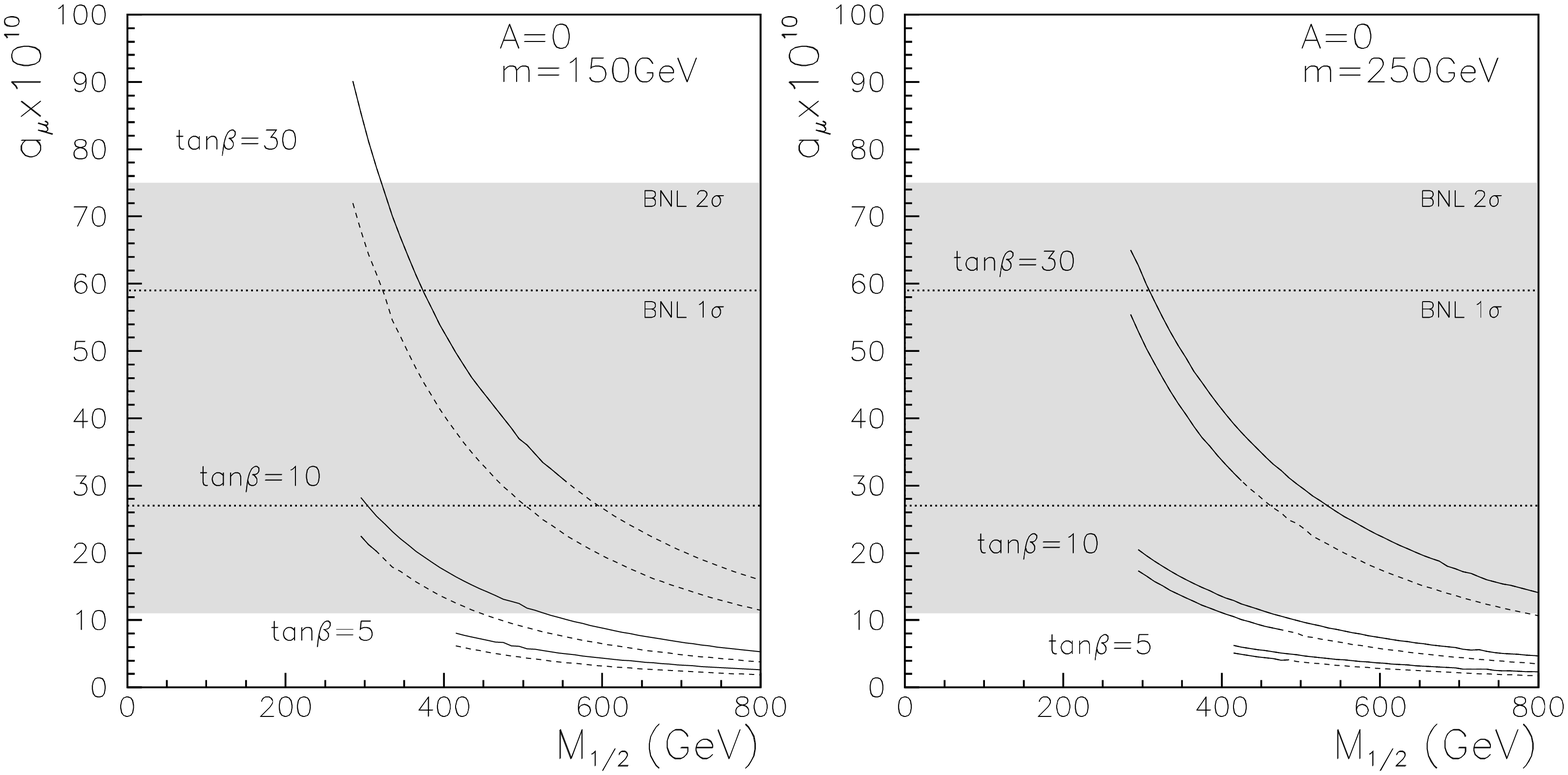, width=15cm, height=7cm}
\end{center}
\vspace{-0.5cm}
\caption{ \asusy\ as a function of the common gaugino mass $M_{1/2}$ for
$M_I=3\times 10^{11}$ GeV.
The two curves inside each set associated
to a particular value of $\tan\beta$ correspond, from bottom to top, to
the cases without and with gauge couplings unification,
respectively.
Continuous lines correspond to regions where the neutralino is the LSP.}
\label{common-m_1/2}
\end{figure}

Let us consider now the case with gauge coupling unification
at $M_I$ mentioned above.
This scenario is analyzed in Fig.~\ref{common-m_1/2} where \asusy\ is plotted 
versus $M_{1/2}$ for $\tgb=5,10,30$, and
$m=150,250$ GeV. 
Each set of curves 
show also the case without gauge 
unification (lower one) studied above for comparison with 
the case with gauge unification (upper one), 
both for $M_I=3\times 10^{11}$ GeV. 

From the plots in Fig.~\ref{common-m_1/2} one can learn that in the case with 
unification the contribution to \asusy\ is increased with respect to the 
case without unification.
The reason being that now $\alpha_2 (M_I)$ is bigger and
therefore the weak gaugino mass $M_2$ is smaller at low energy. 
This is the opposite to what happens in the case of dark matter
analyses.
There, in the case with gauge unification the neutralino-nucleon cross
sections are
decreased.
It is worth remarking that in Fig.~\ref{common-m_1/2} it is the lower bound 
on the Higgs mass, that we set to be $m_{H} \gsim 114$ GeV, which 
prevents the common 
gaugino mass $M_{1/2}$ from taking lower values than about 300 GeV.

\section{$\amu$ in SUSY models with non--universal soft terms}

\subsection{MSSM with non-universality}

As mentioned in the Introduction, the soft SUSY-breaking terms can 
have in general
a non-universal structure in the MSSM. 
Here we will analyze 
the effect induced by this non--universality on \asusy. 
In particular we will see that 
parameterizations of the soft terms producing an enhancement of 
\asusy\ are possible, making it within the experimental
limit even with low $\tan\beta$, unlike the universal cases.

The SUSY 
contributions to \asusy\ depend essentially on the 
gaugino masses $M_i$, the slepton masses 
$m_{l_L}^2$, $m_{e_R}^2$, and the values of $\mu$,
$A_{\mu}$ and $\tan \beta$. As we discussed above, small $\mu$, 
$m_{\tilde{\nu}}$, $M_2$
are favored to enhance \asusy. Therefore, here, we consider an scenario with 
non--universal soft breaking terms at $M_{GUT}$ where the sleptons and Higgs 
masses are parameterized by
\begin{eqnarray}
&& m_{H_2}^2 = a~ m^2~,~~~~~~~~~~~~~~~~~~~~~~~~~~~~~~~a > 1 \;, \nonumber \\
&& m_{H_1}^2 = m_{l_L}^2 = m_{e_R}^2 = b~ m^2 ~,~~~~~~~~0 \leq b < 1
\;.
\label{mimi}
\end{eqnarray}
The squark masses, which are irrelevant for this analysis, are assumed to be 
universal and equal to $m$. 
Since the smaller(bigger) $m_{H_1}^2$($m_{H_2}^2$) at $M_{GUT}$ is,
the less important the positive(negative) contribution
to $\mu$ at the electroweak scale becomes,
the above non-universality for Higgs masses will decrease the value of
$\mu$. Reducing the soft slepton masses we also reduce the
sneutrino and smuon masses.
The gaugino masses are also assumed to 
be non-universal, as we will discuss below,
and we have fixed $M_2$ such that the lightest chargino mass at the 
weak scale is 
of the order of 
the current experimental limit, i.e. $\mathcal{O}(100)$ GeV. 
Finally we assume that the A-terms are vanishing except for $A_{\mu}$.

\begin{figure}[t] 
\begin{center}
\psfig{file=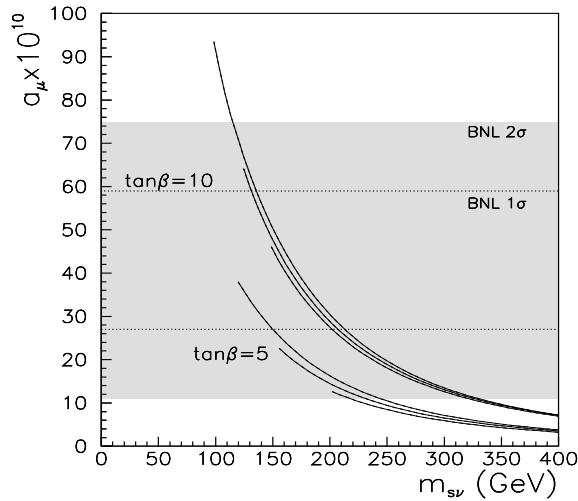, width=8cm, height=7cm}
\end{center}
\vspace{-0.5cm}
\caption{\asusy\ as a function of the sneutrino mass $m_{\tilde{\nu}}$ in
the MSSM 
with the non-universal soft terms discussed in the text.
The three curves inside each set associated to a particular value
of $\tgb$ correspond, from top to bottom, to $A= -3m, 0, 3 m$ 
respectively .}
\label{nonuniversal}
\end{figure}

We find that in this class of models, it is possible to obtain \asusy\ 
within the E821 $1\sigma$ bounds with low $\tan \beta$. 
In Fig.~\ref{nonuniversal}
we present the results for \asusy\ as a function of the sneutrino mass for 
$\tan \beta =5$ and 10. We have assumed that $a=2$, $b= 0.5$ and $m$ varies 
from 150 to 600 GeV. We also fix $M_1 = M_2 = 140$ GeV which leads 
to a lightest
chargino mass of order 120 GeV and a lightest neutralino mass of order 
60 GeV. Also 
we need to take large values for
$M_3$ of order $m$. In this particular example
we use $M_3 = \sqrt{3} m$. Three values for $A_{\mu}$ have 
been examined,
namely $A_{\mu} = -3 m, 0, 3 m$. As we can see from 
Fig.~\ref{nonuniversal}
large and negative values of $A_{\mu}$ allow larger values of \asusy.  

It is worth noticing here that unlike the case with intermediate
scales,
now the neutralino contribution is positive, 
helping in increasing the value of \asusy. 
In any case, still the dominant contribution is due to the chargino.

It is remarkable that the non-universality of 
the soft SUSY-breaking terms has a
very important role in enhancing the values of \asusy\ and 
making it within 
the 
experimental limit even with low $\tan \beta$. In fact, the non-universality 
of the 
gaugino masses is crucial for such a enhancing. In the 
case of universal gaugino 
masses $M_1 = M_2 = M_3 = M_{1/2}$, the lower bound of the Higgs 
mass ($m_{H} \gsim 114$ GeV) requires large values of $M_{1/2}$. 
Recall that the Higgs mass gets a large contribution from the
loop correction which is proportional to the stop mass.
Since
we want to have the lightest 
neutralino as the LSP we then need to push $m$ to higher values 
in order to avoid some slepton as the LSP. This, of course,
leads 
to a heavy spectrum and hence 
\asusy\ is supressed. Relaxing this assumption, we can have $M_3$ large
in order to fulfil the Higgs bound, 
we can keep $M_1$ light to assure that the LSP is always 
the lightest neutralino,
and we can 
have $m$ not very heavy. Furthermore, we found that if one assumes 
non-universality only in the gaugino sector, 
the enhancement 
of \asusy\ is not enough to be in the E821 regions. One still needs
to assume 
that the slepton and Higgs masses are non-universal as 
discussed below eq.(\ref{mimi}).

\subsection{D--brane models}

Recent studies of type I strings have shown that it is possible to construct 
a number of models with  
interesting phenomenological properties \cite{kaku2,typeITeV,otro}. 
It was also shown that
models with the gauge group and particle content of the
supersymmetric standard model lead naturally to 
intermediate values for the
string
scale, in order to reproduce the value of gauge couplings
deduced from experiments \cite{nosotros}. In addition, non-universal
soft SUSY-breaking terms appear generically.

Type I models contain D-branes
and the gauge groups of the SM may come from different
types of D-branes or from the same type of D-branes.
Although, as mentioned above, intermediate values for the string scale
($M_I=10^{10-12}$ GeV)
are naturally obtained we will consider here a model where
also higher values are allowed. This will allow us 
to study 
the variations of $\amu$ as a 
function of the initial scale $M_I$, following the lines of 
Section 3. 

In particular, in this model the gauge group
$U(3)\times U(2)\times U(1)$, giving rise to 
$SU(3)\times SU(2)\times U(1)^3$, arises 
from three different types of D-branes, and therefore
the gauge couplings will be non-universal.
Interesting phenomenological properties of this model can be found
in refs.\cite{nosotros,edm}.
The analysis of the soft terms has been done under 
the 
assumption that only the 
dilaton ($S$) and moduli ($T_i$) fields contribute to SUSY
breaking and it has been found  that these soft terms  are generically 
non-universal.
Using the standard parameterization \cite{bim}
\begin{eqnarray}
&& F^S= \sqrt{3} (S+S^*) m_{3/2} \sin \theta\;, \nonumber \\
&& F^i= \sqrt{3} (T_i+T^*_i) m_{3/2} \cos \theta\; \Theta_i\;,
\end{eqnarray}
where
$i=1,2,3$ labels the three complex compact dimensions, and 
the angle $\theta$ and the $\Theta_i$ with $\sum_{i} |\Theta_i|^2=1$,
just parametrize the direction of the goldstino in the $S$, $T_i$ field
space, one is able to obtain the following soft terms \cite{nosotros}.
The gaugino masses are given by
\bea
M_3 & = & \sqrt{3} m_{3/2} \sin \theta \ , \nn\\
M_{2} & = & \sqrt{3}  m_{3/2}\ \Theta_1 \cos \theta  \ , \nn\\
M_{Y} & = &  \sqrt{3}  m_{3/2}\ \alpha_Y (M_I)
\left(\frac{2\ \Theta_3 \cos \theta}{\alpha_1 (M_I)}
+\frac{\Theta_1 \cos \theta}{\alpha_2 (M_I)}
+\frac{2 \sin \theta}{3 \alpha_3 (M_I)}
\right)\ ,
\label{gaugino1}
\eea
where
\begin{equation}
\frac{1}{\alpha_Y(M_I)} =
\frac{2}{\alpha_1(M_I)} + \frac{1}{\alpha_2(M_I)}
+ \frac{2}{3 \alpha_3(M_I)}\ .
\label{couplings}
\end{equation}  
Here, relation (\ref{couplings}) is due to the D-brane origin of the
$U(1)$ gauge groups. In particular 
$U(1)_Y$ is a linear combination 
of the three $U(1)$ gauge groups arising from $U(3)$, $U(2)$ and
$U(1)$ within three different D-branes.
$\alpha_k$ correspond to the gauge couplings of the $U(k)$ branes.
As  shown in Ref.\cite{nosotros},  $\alpha_1(M_I) = 0.1(1)$ leads
to the string scale $M_I = 10^{12}(5\times 10^{15})$ GeV.

The soft scalar masses are given by
\begin{eqnarray} 
m^2_{q} & = & m_{3/2}^2\left[1 -
\frac{3}{2}  \left(1 - \Theta_{1}^2 \right)
\cos^2 \theta \right] \ , \nn \\
m^2_{d^c} & = & m_{3/2}^2\left[1 -
\frac{3}{2}  \left(1 - \Theta_{2}^2 \right)
\cos^2 \theta \right] \ , \nn \\
m^2_{u^c} & = & m_{3/2}^2\left[1 -
\frac{3}{2}  \left(1 - \Theta_{3}^2 \right)
\cos^2 \theta \right] \ , \nn \\
m^2_{e^c} & = & m_{3/2}^2\left[1- \frac{3}{2}
\left(\sin^2\theta + \Theta_{1}^2 \cos^2\theta  \right)\right] \ , \nn \\
m^2_{l} & = & m_{3/2}^2\left[1- \frac{3}{2}
\left(\sin^2\theta + \Theta_{3}^2 \cos^2\theta  \right)\right] \ , \nn \\
m^2_{H_2} & = & m_{3/2}^2\left[1- \frac{3}{2}
\left(\sin^2\theta + \Theta_{2}^2 \cos^2\theta  \right)\right] \ , \nn \\
m^2_{H_1} & = & m^2_l \;,
\label{scalars1}
\end{eqnarray}
and finally the trilinear parameters are
\begin{eqnarray}
A_{u} & = &  \frac{\sqrt 3}{2}m_{3/2}
   \left[\left(\Theta_{2} - \Theta_1 
 - \Theta _{3}  \right) \cos\theta
- \sin\theta \; \right] \ ,  
\nn \\
A_{d} & = &  \frac{\sqrt 3}{2}m_{3/2}
   \left[\left(\Theta_{3} - \Theta_1 
  - \Theta _{2}  \right) \cos\theta
- \sin\theta \; \right] \ ,
\nn \\
A_{e} & = &  0\; .
\label{trilin11}
\end{eqnarray}

\begin{figure} 
\begin{center}
\psfig{file=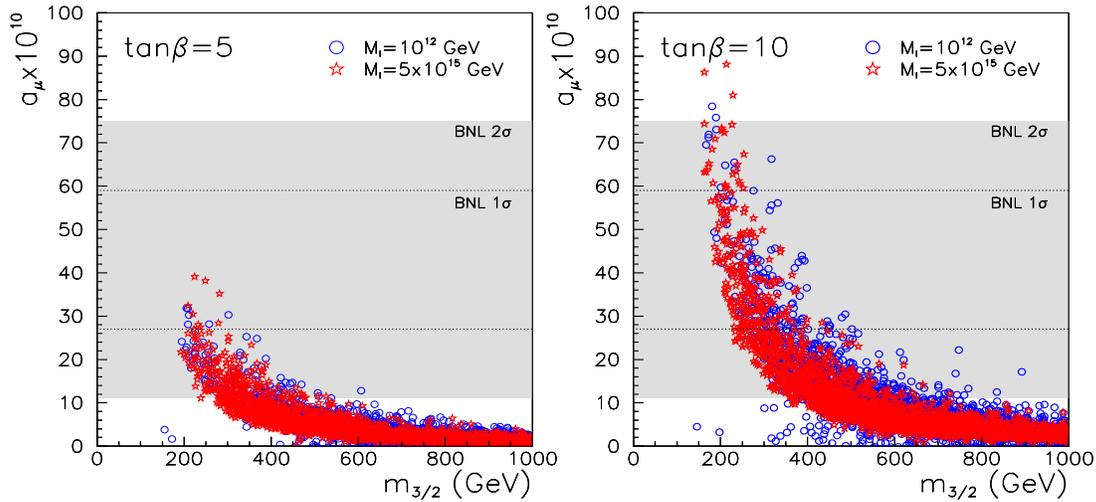, width=15cm, height=7cm}
\end{center}
\vspace{-0.5cm}
\caption{\asusy\ as a function of 
the gravitino mass $m_{3/2}$ in the D-brane model,
for two values of the string scale $M_I= 10^{12}$ and $5\times 10^{15}$
GeV,
and for $\tgb = 5, 10$.}
\label{dbranes}
\end{figure}

We observe that the angle $\theta$ and the $\Theta_i$ 
are quite constrained in order
to avoid negative mass-squared for squarks and sleptons.
This constraint allows a small region for the angle $\theta$, namely
$ 0 < \theta \lsim \pi/6$. 
From eq.(\ref{gaugino1}), one notes that in this allowed
region of $\theta$ we have at the 
string scale $M_3 < M_2 < M_1$. However, at the 
electroweak scale and due to the different running we find that $M_2$ is the 
lightest gaugino mass, namely $M_2 < M_1 < M_3$.
This is an interesting 
example for having 
the lightest chargino mass not very heavy and hence \asusy\ 
could be enhanced as discussed in Section 2.

In Fig.~\ref{dbranes} we show a scatter plot of \asusy\ as a function of the 
gravitino mass $m_{3/2}$ for a scanning of the parameter
space discussed above. Two different values of $\tan \beta$, 5 and 10,
are shown. Likewise, we consider 
the two values of the string scale discussed previously,
$M_I= 10^{12}$ and $5\times 10^{15}$ GeV. 
Clearly,
these models are much more constrained than the generic MSSM with
non-universal soft terms. However, 
we can see 
that for a string scale of order $10^{12}$ GeV 
the values of \asusy are within the 
E821 $1\sigma$ for 
$\tan \beta \gsim 5$. 
As expected from the discussion in Section 3, smaller values of 
\asusy\ are obtained for the scale  $5\times 10^{15}$ GeV.

\section{Conclusions}

In the light of the new BNL results on muon g-2, we have analyzed
\asusy\ in SUSY scenarios.
First we have concentrated on scenarios with universal soft terms.
In particular, we have analyzed
the sensitivity of \asusy\ with respect to
the initial scales $M_I$, smaller than $M_{GUT}$, where
soft SUSY--breaking terms are generated.

We have noted a moderate sensitivity of \asusy\ to the value of the
initial scales for the running of the soft terms. We found that
the smaller the scale is the larger \asusy\ becomes.   
In particular, by taking $M_I\approx 10^{10-12}$ GeV rather than
$M_{GUT}\approx 10^{16}$ GeV, which is a more sensible choice e.g.
in the context of some superstring models, we find that \asusy\ increases
at most of $30 \%$ in the large $\tgb$ region ($\tgb = 30$), 
while it is less than $10 \%$ for $\tgb \le 10$.

We applied
the new BNL results to set upper bounds on the relevant SUSY spectrum,
by requiring that \asusy\ lies within the $2\sigma$ BNL deviation.
The main relevant result of BNL constraints is that
the regions with $\tgb \le 5$ are excluded for any scale
$M_I\approx 10^{10-16}$ GeV.
Besides, we have found upper bounds on the lightest neutralino, chargino,
smuon, and sneutrino masses, namely of the order of
340, 640, 330, 560 GeV respectively, (for $\tgb=30$ and $m=150$) at
$M_I=10^{16}$ GeV. These bounds are increased of ${\cal O}(60 \%)$
in the case of neutralino and are decreased of order ${\cal O}(10-20 \%)$
for the other ones, by decreasing
the initial scale $M_I$ from $10^{16}~{\rm GeV}$ to  $10^{10}$ GeV.

We have also analyzed the corresponding results for the case of
gauge couplings unification at intermediate scale $M_I=10^{11}$ GeV.
In this case we have
found that the values of \asusy\ are higher, at most of 30\%,
with respect to the corresponding ones mentioned above,
with the same initial scale
but without gauge couplings unification.

Then we have studied the possibility of having non-universal soft terms,
which is a generic situation in supergravity and superstring models.
For the usual scale $M_{GUT}\approx 10^{16}$ GeV, we have 
obtained $a_{\mu}$ in the favored experimental range  
even for moderate $\tan\beta$ regions 
($\tan\beta\gsim 5$).
Obviously, from the previous result we can deduce that
lowering the initial scale for the running of the non-universal soft
terms, larges values for \asusy\ will be obtained.

Finally, we have given an explicit example where the two situations
discussed above occur. This is the case of a D-brane model, 
where the string scale is naturally of order $10^{10-12}$ GeV and 
the soft terms are non universal.
We have obtained that  $a_{\mu}$ is enhanced with low $\tan\beta$.

\bigskip

\noindent {\bf Note added}
 
\noindent
As this manuscript was prepared, ref.\cite{Ko} appeared.
The authors discuss also the variation of $a_{\mu}$ with
the initial scale, however, their analysis concentrates on the
dilaton limit.

\bigskip

\noindent {\bf Acknowledgments}

\noindent
D.G. Cerde\~no acknowledges the financial support
of the Comunidad de Madrid through a FPI grant.
The work of
S. Khalil
was supported by PPARC.
The work of C. Mu\~noz was supported
in part by the Ministerio de Ciencia y Tecnolog\'{\i}a,
and
the European Union under contract HPRN-CT-2000-00148.
The work of E. Torrente-Lujan was supported in part by
the Ministerio de Ciencia y Tecnolog\'{\i}a.



\begin{thebibliography}{99}
\bibitem{g2_rev} A. Czarnecki and W.J. Marciano, hep-ph/0102122.

\bibitem{davier} M. Davier, and A. H\"ocker, \PLB{435}{98}{427};
M. Davier, hep-ex/9912044; 
S. Eidelman and F. Jegerlehner, Z. Phys. C67, 585 (1995); K. Adel
and F.J. Yndurain, Rev. Acad. Ciencias (Esp.), 92 (1998), hep-ph/9509378;
J.A. Casas, C. L\'opez and F.J. Yndurain, Phys. Rev. 32, 736 (1985);
S. Narison, hep-ph/0103199.





\bibitem{g2_new}
E.A. Baltz and P. Gondolo, hep-ph0102147;
J. Feng and K.T. Matchev, hep-ph/0102146; L. Everett, G.L. Kane,
S. Rigolin, and L.T. Wang, hep-ph/0102145; T. Ibrahim, U. Chattopadhyay, 
and P. Nath, hep-ph/0102324; 
R. Arnowitt, B. Dutta, B. Hu and Y. Santoso, hep-ph/0102344;
K. Choi, K. Hwang, 
S.K. Kang, K.Y. Lee, and W.Y. Song, hep-ph/0103048;
A. Dedes and H. E. Haber, hep-ph/0102297;
S. Martin and J.D. Wells, hep-ph/0103067; S. Komine, T. Moroi, and
M. Yamaguchi, hep-ph/0103182; H. Baer, C. Bal\'azs, J. Ferrandis, and
X. Tata, hep-ph/0103280; 
D.F. Carvalho, J. Ellis, M.E. Gomez and S. Lola, hep-ph/0103256;
K. Enqvist, E. Gabrielli and K. Huitu, hep-ph/0104174.

\bibitem{g2_exot} D. Chakraverty, D. Choudhury, and A. Datta, hep-ph/0102180;
K. Cheung, hep-ph/0102238;
X. Calmet, H. Fritzsch and D. Holtmannsp\"otter, hep-ph/0103012;
M.B. Einhorn and J. Wudka, hep-ph/0103034;
K. Cheung, C.-H. Chou and O.C.W. Kong hep-ph/0103183.

\bibitem{yndurain} F.J. Yndur\'ain, hep-ph/0102312.

\bibitem{marro} W.J. Marciano and B.L. Roberts, hep-ph/0105056.



\bibitem{nath1} U. Chattopadhyay and P. Nath, hep-ph/0102157.

\bibitem{BNL} H.N. Brownn {\it et al.}, hep-ex/0102017.


\bibitem{reviewdienes} For a review, see: 
K.R. Dienes, Phys. Rept. 287 (1997) 447, and references therein.

\bibitem{Witten} E. Witten, Nucl. Phys. B471 (1996) 135.

\bibitem{Rigolin} L.E. Ibanez, C. Mu\~noz and S. Rigolin, \NPB{553}{99}{43}.

\bibitem{Banks} T. Banks and M. Dine, \NPB{479}{96}{173};
D.G. Cerde\~no and C. Mu\~noz, Phys. Rev. D61 (2000) 016001.

\bibitem{Lykken} J. Lykken, Phys. Rev. D54 (1996) 3693; 
N. Arkani-Hamed, S. Dimopoulos and G. Dvali,
Phys. Rev. Lett. B249 (1998) 262;
I. Antoniadis, N. Arkani-Hamed, S. Dimopoulos and
G. Dvali, Phys. Lett. B436 (1998) 263;
I. Antoniadis and C. Bachas, Phys. Lett. B450 (1999) 83.

\bibitem{typeITeV} G. Shiu and S.-H.H. Tye, Phys. Rev. D58 (1998) 106007;
Z. Kakushadze and S.-H.H. Tye, Phys. Rev. D58 (1998) 126001.

\bibitem{Kiritsis} I. Antoniadis, E. Kiritsis and T.N. Tomaras, 
Phys. Lett. B486 (2000) 186.


\bibitem{typeIinter} C. Burgess, L.E. Iba\~nez and F. Quevedo,
Phys. Lett. B447 (1999) 257.


\bibitem{nosotros} D.G. Cerde\~no, E. Gabrielli, S. Khalil, C. Mu\~noz
and E. Torrente-Lujan, hep-ph/0102270, to appear in Nucl. Phys. B.

\bibitem{typeII} I. Antoniadis and B. Pioline, 
Nucl. Phys. B550 (1999) 41. 

\bibitem{stronghete} K. Benakli, Phys. Rev. D60 (1999) 104002.

\bibitem{weakandstronghete} K. Benakli and Y. Oz, Phys. Lett. B472 
(2000) 83; A. Gregori, hep-th/0005198.

\bibitem{Allanach} 
S.A. Abel, B.C. Allanach, F. Quevedo, L.E. Ib\'a\~nez and M. Klein,
JHEP 0012 (2000) 026.

\bibitem{mua} For a review, see e.g.: C. Mu\~noz, hep-ph/9709329, and
references therein.

\bibitem{caos} N. Kaloper and A. Linde, Phys. Rev. D59 (1999) 101303.

\bibitem{muas} E. Gabrielli, S. Khalil, C. Mu\~noz
and E. Torrente-Lujan, Phys. Rev. D63 (2001) 025008. 

\bibitem{bailin} D. Bailin, G.V. Kraniotis and A. Love,
Phys. Lett. B491 (2000) 161.


\bibitem{experimento1}
R. Bernabei et al., Phys. Lett. B480 (2000) 23.

\bibitem{experimento2}
R. Abusaidi et al., Phys. Rev. Lett. 84 (2000) 5699.

\bibitem{Bottino} A. Bottino, F. Donato, N. Fornengo and S. Scopel, 
\PRD{59}{99}{095004}; R. Arnowitt and P. Nath, \PRD{60}{99}{044004};
E. Accomando, R. Arnowitt, B. Dutta and Y. Santoso, 
Nucl. Phys. B585 (2000) 124.

\bibitem{Gomez}
M.E. G\'omez and J.D. Vergados, hep-ph/0012020.

\bibitem{mirage} L.E. Iba\~nez, hep-ph/9905349;
I. Antoniadis, C. Bachas and E. Dudas, 
Nucl. Phys. B560 (1999) 93;
N. Arkani-Hamed, S. Dimopoulos and J. March-Russell,
hep-th/9908146.

\bibitem{dilaton} 
For a review, see e.g.: A. Brignole, L.E. Ibanez and C. Mu\~noz, in the 
book `Perspectives on Supersymmetry', World Scientific Publ. Co.
(1998) 125, hep-ph/9707209, and references therein.

\bibitem{nath} U. Chattopadhyay and P. Nath, \PRD{53}{96}{1648}.

\bibitem{moroi} T. Moroi, \PRD{53}{96}{6565}; 56, 4424(E) (1997). 

\bibitem{g2_rec} 
M. Carena, G.F. Giudice, and  C.E.M. Wagner, \PLB{390}{97}{234};
U. Chattopadhyay, D.K. Ghosh, and S. Roy, Phys. Rev. D62, 115001
(2000);
K.T. Mahanthappa and S. Oh, Phys. Rev. D62 (2000) 015012.

\bibitem{g2_gs} E. Gabrielli and U. Sarid, \PRL{79}{97}{4752};

\bibitem{tan}
J.L. Lopez, D.V. Nanopoulos and X. Wang, Phys. Rev. D49 (1991) 366.


\bibitem{kaku2} Z. Kakushadze, Phys. Lett. B434 (1998) 269;
J. Lykken, E. Poppitz, S. Trivedi, Nucl. Phys. B543 (1999) 105.

\bibitem{otro}  
G. Aldazabal, L.E. Ib\'a\~nez, F. Quevedo, hep-ph/0001083;
M. Cvetic, M. Pl\"umacher, J. Wang,
JHEP 0004 (2000) 004;
G. Aldazabal, L.E. Ib\'a\~nez, F. Quevedo 
and A.M. Uranga, JHEP 0008 (2000) 002;
D. Bailin, G.V. Kraniotis and A. Love, hep-th/0011289;
S.F. King and D.A.J. Rayner, hep-ph/0012076.




\bibitem{bim} A. Brignole, L.E. Ib\'{a}\~{n}ez and C. Mu\~noz,
Nucl. Phys. B422 (1994) 125, B436 (1995) 747 (E); 
A. Brignole, L.E. Ib\'{a}\~{n}ez, C. Mu\~noz and C.
Scheich, Z. Phys. C74 (1997) 157.


\bibitem{edm} S. Abel, S. Khalil and O. Lebedev, hep-ph/0103031 ;
hep-ph/0103320.

\bibitem{Ko} S. Baek, P. Ko and H.S. Lee,
hep-ph/0103218.

\end{thebibliography}
\end{document}